\documentclass{article}
\usepackage{amsfonts}
\usepackage{amsmath}

\newenvironment{proof}[1][Proof]{\noindent\textbf{#1.} }{\ \rule{0.5em}{0.5em}}
\input{tcilatex}

\begin{document}

\begin{center}
{\LARGE Two Cellular Automata for the 3x+1\ Map\bigskip \bigskip }

\textbf{M. Bruschi\smallskip }

Dipartimento di Fisica, Universit\`{a} di Roma "La Sapienza", 00185 Roma,
Italy

Istituto Nazionale di Fisica Nucleare, Sezione di Roma

\textit{e-mail:}\emph{Mario.Bruschi@roma1.infn.it\bigskip }

\bigskip \bigskip

\textit{Abstract}
\end{center}

Two simple Cellular Automata, which mimic the Collatz-Ulam iterated map ($%
3x+1$ map), are introduced. These Cellular Automata allow to test
efficiently the Collatz conjecture for very large numbers.\bigskip

\bigskip \bigskip


\textit{\newpage }

\section{Introduction}

Cellular Automata (CA) were first introduced in the early fifties by J. Von
Neumann \cite{Von N} in his investigation of "complexity", following an
inspired suggestion by S. Ulam. S. Ulam himself was among the first to study
intensively CA as well the nice properties of simple iterated maps (among
them the $3x+1$ map and the related Collatz conjecture, see below)\cite{ulam}%
. Likely, he should been amused to know that there exist simple CA which
"compute" this map. Unfortunately, only a very faint light seems to arise
from these CA concerning the proof of the Collatz conjecture.\medskip

\section{The 3x+1 map and the related conjecture}

The $3x+1$ iterated map was introduced in 1937 by L. Collatz, it was
investigated by a lot of people and it is also known as Collatz map
(sequence), Hasse's algorithm, Syracuse algorithm; the related conjecture is
known as $3x+1$ problem, Collatz conjecture, Ulam problem, Kakutani's
problem, Syracuse problem, Thwaites conjecture (a large literature on this
map, on the related conjecture and on possible generalizations is available,
see [3-16]).\smallskip

\emph{The Collatz map (CM):\smallskip }

\begin{subequations}
\label{cv}
\begin{equation}
u(t+1)=3u(t)+1\ \ \ if\ \ u(t)\ is\ odd;\   \label{cm1}
\end{equation}

\begin{equation}
u(t+1)=\frac{u(t)}{2}\ \ \ if\ \ u(t)\ is\ even;  \label{cm2}
\end{equation}%
\begin{equation}
t\in 
\mathbb{N}
,\ u(0)\ \func{positive}\ \func{integer}.  \label{cm3}
\end{equation}

\emph{The Collatz conjecture (CC):\smallskip } 
\end{subequations}
\begin{equation}
For\ any\ initial\ u(0),the\ CM\ attains\ the\ final\ cycle\ ...1,4,2,1\ ...
\label{cc}
\end{equation}%
\medskip

Surprisingly enough, the number of steps required to attain the final cycle
varies sensibly (and apparently unpredictably) also for contiguous small
initial numbers: f.i.

$u(0)=6\rightarrow 3,10,5,16,8,4,2,1...;\ $

$u(0)=7\rightarrow 22,11,34,17,52,26,13,40,20,10,5,16,8,4,2,1...;$

$\ u(0)=8\rightarrow 4,2,1...;$

thus respectively $8,16,3$ steps are needed. Of course numerical
computations of CM always confirmed CC, but, in spite of the simplicity of
the CM itself, no proof (or disproof) of the CC is yet available. Lagarias
(1985) showed that there are no nontrivial cycles with length \TEXTsymbol{<}%
275000 [3,4]. Conway (1972) proved that Collatz-type problems can be
formally undecidable [5]. Thwaites (1996) has offered a \pounds 1000 reward
for resolving the conjecture \cite{Tw}. \medskip

\section{Two CA that "compute" the Collatz map}

Let us now introduce two simple Cellular Automata that "mimic" the CM.
Looking at the CM itself (\ref{cv}) it is clear that computations should be
very simple in basis $2$ or$\ 3$: we take advantage of this. Indeed the
first automaton (CA2) computes the map in basis $2$, while the second one
(CA3) computes it in basis $3$.\bigskip

\subsection{CA2}

This is an unidimensional CA, the cells are arranged on a line, they can be
numbered by $n\in 
\mathbb{Z}
$ and each cell at the discrete time $t$ ($t=0,1,2,...)$ can be in one of
two different states so that the state-function $u(n,t)$ takes values in a
finite set: $u(n,t)\in 
\mathbb{Z}
/2%
\mathbb{Z}
$ , say $\left\{ 0,1\right\} .$The \emph{vacuum-state} is $0,$ i.e. 
\begin{equation}
u(n,t)\longrightarrow 0\ \ as\ \ \left\vert n\right\vert \longrightarrow
\infty .  \label{vac}
\end{equation}

\textit{Example:}

$...00011010000111010001011001000...\ $

The \emph{evolution law}, that allows to construct the state $u(n,t+1)$ of
the CA from the known $u(n,t)$, consists in the following rules:\bigskip

\textsc{rule Ia}: running from the left to the right (increasing $n$) tag
the cells from the first $0$ followed by a $1$ to the first $0$ followed by $%
0$ (extrema included);\smallskip

\textit{applying this rule} \textit{in the above example, and tagging with a
superimposed dot, we get:\medskip }

$...00\dot{0}\dot{1}\dot{1}\dot{0}\dot{1}\dot{0}000111010001011001000...%
\bigskip $

\textsc{rule Ib: }tag also, if the case, from the next $1$ following a $1$
to the first $0$ followed by $0$ (extrema included);

\textit{applying this rule} \textit{in the above example we get:\medskip }

$...00\dot{0}\dot{1}\dot{1}\dot{0}\dot{1}\dot{0}0001\dot{1}\dot{1}\dot{0}%
\dot{1}\dot{0}001011001000...\bigskip $

\textsc{rule Ic: }proceeding to the right, apply \textsc{rule Ib }whenever
possible:\smallskip

\textit{applying this rule} \textit{in the above example we get:\medskip }

$...00\dot{0}\dot{1}\dot{1}\dot{0}\dot{1}\dot{0}0001\dot{1}\dot{1}\dot{0}%
\dot{1}\dot{0}00101\dot{1}\dot{0}01000...\bigskip $

\bigskip \textsc{rule II:} the evolved state obtains according to:\smallskip 
\begin{subequations}
\label{r2}
\begin{equation}
u(n-1,t+1)=\left( u(n,t)+u(n+1,t)\right) \ \func{mod}\ 2,\ \ if\ the\ cell\
n\ is\ not\ tagged;  \label{r2a}
\end{equation}

\begin{equation}
u(n-1,t+1)=\left( u(n,t)+u(n+1,t)+1\right) \ \func{mod}\ 2,\ \ if\ the\
cell\ n\ is\ tagged.  \label{r2b}
\end{equation}

\smallskip \textit{Applying this rule} \textit{in the above example we
get:\medskip }

$...00100010010100010111001110000...\bigskip $

Let us now show how this CA computes the Collatz map. Given a CA2
configuration $u(n,t),$ f.i.:

$u(n,t)=\ ...00010011000...,$

consider the configuration $\tilde{u}(n,t)$ which goes from the leftmost $1$
to the rightmost $1$:

$\tilde{u}(n,t)=\ 10011.$

A configuration $\tilde{u}(n,t)$ is initially finite, due to (\ref{vac}),
and stays finite during the evolution, due to rules I and II. Thus we can
interpret $\tilde{u}(n,t)$ as an integer positive (odd) number in basis $2$
(proceeding from the left to the right):

$\tilde{u}(n,t)=\ 10011=\ 1\ast 2^{0}+0\ast 2^{1}+0\ast 2^{2}+1\ast
2^{3}+1\ast 2^{4}=25$

It is straightforward to see that, considering the configurations $\tilde{u}%
(n,t)$ as positive (odd) integers, CA2 emulates (\ref{cm1}).

\begin{proof}
To multiply a number by $3$ (namely $11$ in basis $2$), we have to perform
the binary summation of that number and of the same number times $2$ (which
in our CA2 configuration is the number itself shifted to the right). In
other words we have to add any digit to the following one, taking care of
the possible amount to be carried: indeed we have to carry $\ 1$ after a $1$
following a $1$ up to the first $0$ following a $0$. This justifies Rules
Ib, Ic, and II. Rule Ia just adds $1$ to the result.
\end{proof}

\bigskip

Now note that the CM rule (\ref{cm2}) in our framework is just a trivial
shift of the configuration. Thus we can assert that this CA emulate the
Collatz map. The Collatz conjecture (\ref{cc}) in the language of this CA
reads:

"\textit{any initial (non zero) configuration evolves attaining eventually
the \emph{stable} configuration }$...0001000...$"

\bigskip

Let us give a simple example:

$u(n,0)=\ ...00010011000...\ $

$u(n,1)=\ ...00011001000...\ $

$u(n,2)=\ ...00101110000...\ $

$u(n,3)=\ ...00011010000...\ $

$u(n,4)=\ ...00100010000...\ $

$u(n,5)=\ ...00101100000...\ $

$u(n,6)=\ ...00010100000...\ $

$u(n,7)=\ ...00000100000...\ $

Interpreting now the configurations as integers, we have the sequence:

\textbf{25,19,29,11,17,13,5,1...}

which is the \emph{odd }subsequence of the Collatz sequence:

\textbf{25},76,38,\textbf{19},58,\textbf{29},88,44,22,\textbf{11},34,\textbf{%
17},52,\textbf{13},40,20,10,\textbf{5},16,8,4,2,\textbf{1}...

Note the efficiency of CA2 which reaches the final states in only $7$ steps
(versus the $23$ steps needed by the Collatz map). The typical behavior of
this automaton is shown in Figure 1.

\subsection{CA3}

This is an unidimensional CA, the cells are arranged on a line, they can be
numbered by $n\in 
\mathbb{Z}
$ and each cell at the discrete time $t$ ($t=0,1,2,...)$ can be in one of
four different states so that the state-function $u(n,t)$ takes values in a
finite set: $u(n,t)\in 
\mathbb{Z}
/2%
\mathbb{Z}
$ , say $\left\{ 0,1,2,3\right\} .$Moreover 
\end{subequations}
\begin{subequations}
\label{c3}
\begin{equation}
u(n,t)\underset{n\rightarrow -\infty }{\longrightarrow }0\ ;  \label{c3a}
\end{equation}%
\begin{equation}
u(n,t)\underset{n\rightarrow +\infty }{\longrightarrow }3;  \label{c3bis}
\end{equation}%
\begin{equation}
If\ \ u(n,t)=3\ \ then\ \ u(n+1,t)=3.  \label{c3ter}
\end{equation}

\textit{Example:}

$...00012010222102110012333...$

The \emph{evolution law}, that allows to construct the state $u(n,t+1)$ of
the CA from the known $u(n,t)$, consists in the following two rules:\bigskip

\textsc{rule III:} tag the cells following a non tagged $1$ up to the next $%
1 $ or $3$ (included);

\textit{applying this rule} \textit{in the above example, and tagging with a
superimposed dot, we get:\medskip }

$...0001\dot{2}\dot{0}\dot{1}02221\dot{0}\dot{2}\dot{1}01\dot{1}001\dot{2}%
\dot{3}33...\medskip $

\textsc{rule IV:} the evolved state of CA3 obtains according to the
following table:\smallskip \smallskip 
\end{subequations}
\begin{equation}
\begin{tabular}{cc}
tagged cells & not tagged cells \\ 
$0\rightarrow 1$ & $0\rightarrow 0$ \\ 
$1\rightarrow 2$ & $1\rightarrow 0$ \\ 
$2\rightarrow 2$ & $2\rightarrow 1$ \\ 
$\ 3\rightarrow 2\ $ & $3\rightarrow 3$%
\end{tabular}
\label{tab}
\end{equation}

\textit{applying this rule} \textit{in the above example we get:\medskip }

$...0000212011101220020002233...\medskip $

Let us now show how this CA computes the Collatz map. Given a $u(n,t),$ f.i.:

$u(n,t)=\ ...000101121333...,$

consider the configuration $\tilde{u}(n,t)$ which goes from the leftmost
not-zero number (included) to the first $3$ (excluded):

$\tilde{u}(n,t)=\ 101121.$

A configuration $\tilde{u}(n,t)$ is initially finite, due to (\ref{c3}), and
stays finite during the evolution, due to rules III and IV. Thus we can
interpret $\tilde{u}(n,t)$ as an integer positive number in basis $3$ \
(proceeding from the right to the left):

$\tilde{u}(n,t)=\ 101121$ (basis $3$)$=\ 1\ast 3^{0}+2\ast 3^{1}+1\ast
3^{2}+1\ast 3^{3}+0\ast 3^{4}+1\ast 3^{5}=286$ (basis $10$).

Suppose now that $\tilde{u}(n,t)$ is even (as in the above case); then in $%
\tilde{u}(n,t)$ there is an even number of $1$ and therefore, according to
Rule III, in $u(n,t)$ the first $\ 3$ cannot be tagged:

$u(n,t)=\ ...0001\dot{0}\dot{1}1\dot{2}\dot{1}333...,$

In this case Rules III and IV perform just the division of the number $%
\tilde{u}(n,t)$ by $\ 2$ , according to the CM rule (\ref{cm2}). The proof
of this statement is straight forward and it is based mainly on the trivial
fact that 
\begin{equation}
3^{k+1}=(2+1)\ast 3^{k}.
\end{equation}

Let us explicit this using the above example:

$\tilde{u}(n,t)=\ 101121=1\ast 3^{5}\ +0\ast 3^{4}+1\ast 3^{3}+1\ast
3^{2}+2\ast 3^{1}+1\ast 3^{0}=$

$(2+1)\ast 3^{4}+0\ast 3^{4}+1\ast 3^{3}+1\ast 3^{2}+2\ast 3^{1}+1\ast
3^{0}= $

$2\ast 3^{4}+(2+1)\ast 3^{3}+1\ast 3^{3}+1\ast 3^{2}+2\ast 3^{1}+1\ast
3^{0}= $

$2\ast 3^{4}+4\ast 3^{3}+(2+1)\ast 3^{1}+2\ast 3^{1}+1\ast 3^{0}=$

$2\ast 3^{4}+4\ast 3^{3}+0\ast 3^{2}+4\ast 3^{1}+(2+1)\ast 3^{0}+1\ast
3^{0}= $

$2\ast 3^{4}+4\ast 3^{3}+0\ast 3^{2}+4\ast 3^{1}+4\ast 3^{0}.$

Thus clearly:

$\tilde{u}(n,t)/2=\frac{101121}{2}$ (basis $3$)=$12022\ $(basis $3$)$=143~$%
(basis $10$).

Indeed the tagging Rule III and the Rule IV perform this task on the CA
configuration :

$u(n,t)=\ ...000101121333...;$

apply Rule III:

$\ ...0001\dot{0}\dot{1}1\dot{2}\dot{1}333...,$

apply Rule IV:

$u(n,t+1)=\ ...000012022333...\Longrightarrow \tilde{u}(n,t+1)=12022.$

Let us now consider the case of odd $\tilde{u}(n,t)$, f.i.

$u(n,t)=\ ...000101120333...\ \Longrightarrow \tilde{u}(n,t)=101120\ $(basis 
$3$)=$285\ $(basis $10$).

CM rule (\ref{cm1}) yields

$285\ast 3+1\ $(basis $10$)=$856\ $(basis $10$)=$1011201\ $(basis $3$).

It is plain that in terms of the CA3 configuration this could be
accomplished just by changing the leftmost $\ 3$ into a $\ 1$. But then the
new number is even and thus one should again apply CM rule (\ref{cm2})
yielding as a result:

$285\Longrightarrow 856\Longrightarrow 428\ $(basis $10$)=$120212\ $(basis $%
3 $).

It is straightforward to see that Rules III and IV perform these two
consecutive tasks in just one step.

In the above example:

$\ ...000101120333...\overset{Rule\ III}{\Longrightarrow }\ ...0001\dot{0}%
\dot{1}1\dot{2}\dot{0}\dot{3}33...\overset{Rule\ IV}{\Longrightarrow }\
...000012021233...$

Note that, even if all "computations" are in basis $3$, we need four states
for CA3. Another oddity, in terms of usual unidimensional CA, is the
presence of two different \emph{vacuum states} ($0$ on the left, $3$ on the
right).

Note also that the final cycle, eventually attained according to the Collatz
conjecture, is visualized by CA3 in the following way:

$...0001333...\overset{Rule\ III}{\Longrightarrow }\ ...0001\dot{3}33...%
\overset{Rule\ IV}{\Longrightarrow }\ ...0000233...\overset{Rule\ III}{%
\Longrightarrow }\ ...0000233...\overset{Rule\ IV}{\Longrightarrow }\
...0000133...$

A typical behavior of this CA is shown in Figure 2.

\section{Final remarks}

Looking at Figures 1 and 2, we can see that one edge of the configuration
moves with constant average velocity while the other exhibit erratic
behavior: this somehow justify the conjecture ( obviously prove not!).
However if the introduction of these two CA doesn't seem to give a great
contribution to the aim of proving the Collatz conjecture, on the other hand
it allows to explore intensively the $3x+1$ map also for very large initial
integers. Thus our CA offer a good example of how CA themselves can be used
as fast computing machines even if implemented on sequential computers (much
more if running on dedicated parallel machines). Moreover this approach
allows convenient investigation of the statistical behavior of iterates.
Indeed in \cite{quasi}, the authors proposed a Quasi Cellular Automaton for
the $3x+1$ map: they did not introduce a real CA but merely displayed the
iterated of CM in basis 2, simulating in fact our CA2. In the same work
generic behaviors associated with classes of seed values and periodic and
chaotic structures within iterate patterns were investigated. The same
investigation could be pursued for our CA3.

Finally some further remarks:

\begin{itemize}
\item our CA, according to the Wolfram heuristic classification \cite{wolf},
should be Class 1 CA (thus  they should be not interesting ones...);

\item our CA, due to the tagging rules I and III, are "non local" CA (namely
the evolved state of a cell depends in principle on the whole previous
configuration). Indeed the evolution laws for these CA exhibit some
analogies with the so called "fast rule" introduced in \cite{fast} for the
parity-rule "filter" CA \cite{pr}: this suggests that it should be possible
(may be useful) to find a "local", possibly "filter", formulation of these
CA;

\item empirically, computer investigation shows that "small deformations" of
the evolution law of CA2 and, to a lesser extent, of CA3 give rise mostly to
new CA which themselves end up with final cycles (see Figures 3,4,5: of
course this is not equivalent to change "slightly" the rules of the Collatz
map, thus we have non conventional generalizations of the map itself);

\item the scheme here introduced could be easily and profitably extended to
emulate and explore other different iterated maps.\bigskip 
\end{itemize}

\bigskip {\huge Figure captions}

\textit{Figures 1-5 are space-time diagrams of the evolution of the involved
CA, space on the vertical axis, time flows left-right on the horizontal axis,%
} \textit{the states of the cells are shown in different shades of
grey.\medskip }

\begin{itemize}
\item Figure 1: the typical behavior of CA2, starting from a random chosen
initial configuration of 300 cells (corresponding to an integer of order 2%
\symbol{94}300) and ending with the final cycle predicted by the Collatz
conjecture, is shown.

\item Figure 2: the typical behavior of CA3, starting from a random chosen
initial configuration of 100 cells (corresponding to an integer of order 3%
\symbol{94}100) and ending with the final cycle predicted by the Collatz
conjecture, is shown. Note that, in order to get a final stable
configuration, the average drift velocity was corrected, adding a downward
shift at any 2 time-steps

\item Figure 3: the typical behavior of a variant of CA2, starting from a
random chosen initial configuration of 450 cells (corresponding to an
integer of order 2\symbol{94}450) and ending with the same final cycle of
CA2, is shown. The evolution law of this CA is the same of CA2 except for
the Rule Ib that now reads: \emph{tag also, if the case, from the next }$1$%
\emph{\ following a }$1$\emph{\ or followed by a }$1$\emph{\ to the first }$%
0 $\emph{\ followed by }$0$\emph{\ (extrema included).}

\item Figure 4: the typical behavior of a variant of CA2, starting from a
random chosen initial configuration of 400 cells (corresponding to an
integer of order 2\symbol{94}400) and ending with a final cycle (more
complex than that of CA2), is shown. The evolution law of this CA is the
same of CA2 except for the ending condition of tagging in Rule Ia, Ib that
now reads:\emph{... to the first }$0$\emph{\ followed by }$0$\emph{\ and
following a }$0$\emph{\ (extrema included).}

\item Figure 5: the typical behavior of a variant of CA3, starting from a
random chosen initial configuration of 200 cells (corresponding to an
integer of order 3\symbol{94}200) and ending with a simple final cycle, is
shown. The evolution law of this CA is the same of CA3 except for Rule IV,
that now reads: \emph{tag the cells following a non tagged }$1$\emph{\ up to
the cell that is }$1$\emph{\ or }$3$\emph{\ or is followed by a }$2$\emph{.}%
\newpage 
\end{itemize}

\end{document}